\documentclass[twocolumn,showpacs,preprintnumbers,amsmath,amssymb]{revtex4}  
 
\usepackage{graphicx}  
\usepackage{dcolumn}  
\usepackage{bm}  
 
\begin{document}   
 
\preprint{}  
\input{epsf.tex}  
 
\epsfverbosetrue 
 
\title{Exciton-Polariton scattering for defect detection in cold atom Optical Lattices} 
\author{Hashem Zoubi, and Helmut Ritsch} 
 
\affiliation{Institut f\"{u}r Theoretische Physik, Universit\"{a}t Innsbruck, Technikerstrasse 25, A-6020 Innsbruck, Austria}   
 
\date{29 October, 2007}  
 
\begin{abstract} 
We study the effect of defects in the Mott insulator phase of ultracold atoms in an optical lattice on the dynamics of resonant excitations. Defects, which can either be empty sites in a Mott insulator state with one atom per site or a singly occupied site for a filling factor two, change the dynamics of Frenkel excitons and cavity polaritons. While the vacancies in first case behave like hard sphere scatters for excitons, singly occupied sites in the latter case can lead to attractive or repulsive scattering potentials. We suggest cavity polaritons as observation tool of such defects, and show how the scattering can be controlled in changing the exciton-photon detuning. In the case of asymmetric optical lattice sites we present how the scattering effective potential can be detuned by the cavity photon polarization direction, with the possibility of a crossover from a repulsive into an attractive potential. 
\end{abstract} 
 
\pacs{42.50.-p, 71.36.+c, 71.35.Lk} 
 
\maketitle 
 
\section{Introduction} 

In one of the most significant achievements of cold atom physics in recent years, a degenerate gas of ultracold atoms loaded into an optical lattice formed far off resonance lasers was demonstrated to undergo a quantum phase transition from a superfluid into the Mott insulator phase \cite{Bloch}. In the Mott insulator a perfectly regular lattice at half the laser wavelength with a fixed number of atoms per site is achieved. This quantum phase transition from the superfluid into the Mott insulator was predicted and described by the Bose-Hubbard model \cite{Jaksch,Fisher}.  As the Mott insulator phase can be considered as an artificial crystal with parameters well controllable in time and space, this generates a close connection between cold atom optics and solid state physics, which has generated a plethora of subsequent work generalizing the model and identifying more and more solid state phenomena to be studied in such configurations \cite{Zoller}. Furthermore, strong resonant light-matter coupling for a Bose-Einstein Condensate in an optical cavity has been achieved recently \cite{Esslinger}.
 
In previous work \cite{ZoubiA} we exhibited the similarities of such artificial crystals and molecular or noble atom crystals, where one gets the so called Frenkel-excitons. In particular for the special case of identical ground and excited state optical lattice potentials the Mott insulator phase allows the formation of such excitons as collective electronic excitations in the whole optical lattice \cite{Zoubi,Agranovich}. If the optical lattice is placed between cavity mirrors with a single cavity mode close to resonance with the excitons, coherent superpositions of excitons and cavity photons can form. In the strong coupling regime the corresponding system eigenstates then are cavity polaritons, which have very similar behavior in solid state and cold atom systems \cite{Zoubi,Kavokin}. Going beyond the analogous solid state case, very interesting new physics appears in the case of lattices with two atoms per site, where the on-site resonant excitations form symmetric (bright) and antisymmetric (dark) superposition states of individual excitations \cite{ZoubiB}. While the symmetric states radiate and via dipole-dipole coupling will form excitons and thus cavity polaritons, the antisymmetric states have no dipole moment, so that their excitations are long lived and stay localized at one site. Lattice asymmetries can lead to coupling between bright and dark states and produce strongly polarization dependent spectra. 
 
Optical lattices are spatially and externally described by the laser field and thus almost perfect over long distances. Nevertheless due to imperfections in the dynamical formation of the Mott insulator, the appearance of some defects in the Mott insulator phase is unavoidable \cite{Bloch} and there are missing or extra atoms at some sites. Such defects might create decisive errors when using such lattices for quantum information processing or for observing even more exotic quantum phases in such systems and some suggestion to repair such defects were already presented \cite{Rabl}. On the other hand they are hard to detect by direct observations of the atomic distribution. 
 
In this paper we investigate the effect of such defects on the dynamical properties of excitons and cavity polaritons. We concentrate in the case of a low defect number, where the exciton and cavity polariton picture still holds and these quasi-particles are only scattered by single defects. Namely, the distance between each two defects is enough large for the formation of coherent excitons and cavity polaritons, which propagate as free quasi-particles between each two scattering processes. In principle defects can move by hopping among the lattice sites, but the corresponding time scale is so long that they can be considered frozen. Introducing more and more defects with faster hopping would then correspond to the transition from the Mott to a superfluid atomic phase in the lattice, which will not be considered here.  
 
Here we treat defects related to a missing atom: in the case of one atom per site this corresponds to vacant sites and while for two atoms per site a defect is a singly occupied site. Interestingly in calculating the exciton and cavity polariton elastic scattering amplitude of such defects, the two case behave even qualitatively very different. As a consequence of this defect scattering we show how cavity polaritons can be used as observation tool to detect defects in the Mott insulator phase. 
 
The paper is organized as follows. In section 2 we investigate the exciton scattering off a vacancy in an optical lattice in the Mott insulator with filling factor one, which is generalized to the scattering of cavity polaritons in section 3. The scattering of excitons and cavity polaritons off a defect in the Mott insulator case with two atoms per site is treated in section 4, where also we treat the special case of an optical lattice with asymmetric sites. 
 
\section{Exciton scattering of a vacancy in an optical lattice} 
 
Let us consider first a $2D$ optical lattice at the magic wavelength filled with one 2-level atom per site in the Mott-insulator phase. Including dipole-dipole interactions resonant excitations of the atoms can be represented as quasi-particles called Frenkel excitons \cite{ZoubiA}. A single atom missing at the origin (${\bf r}_i={\bf 0}$) then creates an impurity in the artificial lattice of ultracold atoms, which for sufficient lattice depth stays localized and will not hop among the lattice sites as shown in figure (1). We will now consider scattering of these excitons at such a missing atom (hole). As the scattering time is much shorter than the hopping time, the impurity assumed to be localized during the scattering process.  
 
Adding the impurity to the ideal atom crystal just needs a small change in the system Hamiltonian $H=H_0+V$ derived in Ref.\cite{ZoubiA}, whose components now read: 
\begin{equation} 
H_0=\sum_{\bf k}\hbar\omega_{ex}(k)\ B_{\bf k}^{\dagger}B_{\bf k}\ ,\ V=-\hbar\omega_A\ B_0^{\dagger}B_0. 
\end{equation} 
$H_0$ represents free excitons due to the resonant electronic excitation transfer among lattice sites induced by dipole-dipole interactions including the lattice symmetry. $B_{\bf k}^{\dagger}$ and $B_{\bf k}$ are the creation and annihilation operators of an exciton with in-plane wave vector ${\bf k}$, respectively, and $\omega_{ex}(k)$ is the exciton dispersion. $V$ is the impurity Hamiltonian at the origin, where $\hbar\omega_A$ is the atomic transition energy. 
 
For an exciton the impurity thus appears as a potential well located at the origin, with depth $\hbar\omega_A$ and radius $a$. Here we will not try to find the self-consistent eigenstates of $H$ but to consider only the scattering problem. We can neglect trapping of an exciton in the impurity potential as a vacant site cannot absorb the trapping energy. Hence we consider only the scattering process of an exciton off the impurity. 
 
\begin{figure}[h!] 
\centerline{\epsfxsize=8.0cm \epsfbox{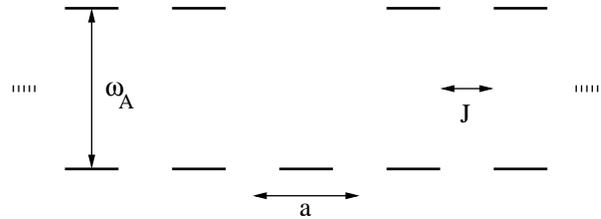}} 
\caption{An impurity in $1D$ optical lattice.} 
\end{figure} 
 
In the following we calculate the scattering amplitude of an exciton of an impurity. The initial exciton has wave vector ${\bf k}$ very far from the impurity, is scattered elastically and found with wave vector ${\bf k'}$ also very far, and for elastic scattering we have $|{\bf k'}|=|{\bf k}|$. As the impurity is very deep with depth $\hbar\omega_A$,  in order to calculate the scattering amplitude we need to go beyond the Born approximation \cite{Taylor} even though the perturbation is confined to a radius $a$.  
 
The eigenstates of the free exciton Hamiltonian are $H_0|\phi_{\bf k}\rangle=E(k)|\phi_{\bf k}\rangle$, and the full Hamiltonian eigenstates obey $H|\psi\rangle=E|\psi\rangle$. The excitation eigenstates in the quasi-momentum space $|\phi_{\bf k}\rangle$ are related to the lattice space eigenstates by $|\phi_{\bf k}\rangle=\frac{1}{\sqrt{N}}\sum_ie^{i{\bf k}\cdot{\bf r}_i}\ |\phi_i\rangle$, where $N$ is the number of lattice sites. The eigenstates form a complete basis, with the closure relation $\sum_i|\phi_i\rangle\langle\phi_i|=\hat{1}$. A general state $|\psi\rangle$ can be expanded by $|\psi\rangle=\sum_i\psi_i\ |\phi_i\rangle$, with the expansion amplitude $\psi_i=\langle\phi_i|\psi\rangle$. 
 
In the scattering problem the incoming exciton is prepared in a delocalized unperturbed exciton state $|\phi_{\bf k}\rangle$ and  the scattered exciton is observed also very far from the impurity. Hence, the whole system eigenstates $|\psi\rangle$ needs to obey the boundary condition $|\psi\rangle\rightarrow|\phi_{\bf k}\rangle$ as $V\rightarrow 0$. The required solution is given by the Schwinger-Lippmann equation \cite{Taylor} 
\begin{equation} 
|\psi\rangle=|\phi_{\bf k}\rangle+G_0V|\psi\rangle, 
\end{equation} 
where 
\begin{equation} \label{Green} 
G_0=\lim_{\eta\rightarrow 0_+}\left(\frac{1}{E-H_0+i\eta}\right). 
\end{equation} 
The sign of $+i\eta$ is chosen in such a way to ensure that the scattered exciton propagates away from the impurity. In multiplying the equation from the left by the bra vector $\langle\phi_i|$, and in inserting the above identity operator between $G_0$ and $V$, we get 
\begin{equation} 
\langle\phi_i|\psi\rangle=\langle\phi_i|\phi_{\bf k}\rangle+\sum_j\langle\phi_i|G_0|\phi_j\rangle\langle\phi_j|V|\psi\rangle. 
\end{equation} 
The calculations using the above definitions yields: 
\begin{equation} 
\psi_i=\frac{1}{\sqrt{N}}e^{i{\bf k}\cdot{\bf r}_i}-\frac{E_a}{N}\sum_{\bf k'}\lim_{\eta\rightarrow 0_+}\frac{e^{i{\bf k'}\cdot{\bf r}_i}}{E-E(k')+i\eta}\ \psi_0, 
\end{equation} 
where $E_a=\hbar\omega_A$. This represents the scattered wave amplitude at site $i$ very far from the impurity. The first term is for the non-scattered wave, and the second is for the scattered wave. For the impurity site amplitude, at ${\bf r}_0={\bf 0}$, we get 
\begin{equation} 
\psi_0=\frac{1}{\sqrt{N}}\left\{1+\frac{E_a}{N}\sum_{\bf k'}\lim_{\eta\rightarrow 0_+}\frac{1}{E-E(k')+i\eta}\right\}^{-1}. 
\end{equation} 
As the scattering is elastic the energy $E$ can be replaced by the incident exciton energy and we have $E=E(k)$. We get 
\begin{equation} 
\psi_i=\frac{1}{\sqrt{N}}\left\{e^{i{\bf k}\cdot{\bf r}_i}+\frac{E_aI_{os}}{1-E_aI_{st}}\right\}, 
\end{equation} 
where we defined the oscillating and static summations 
\begin{eqnarray} 
I_{os}&=&-\frac{1}{N}\sum_{\bf k'}\lim_{\eta\rightarrow 0_+}\frac{e^{i{\bf k'}\cdot{\bf r}_i}}{E(k)-E(k')+i\eta}, \nonumber \\ 
I_{st}&=&-\frac{1}{N}\sum_{\bf k'}\lim_{\eta\rightarrow 0_+}\frac{1}{E(k)-E(k')+i\eta}. 
\end{eqnarray} 
Our main aim now is to calculate the two sums in the case of small wave vector scattering excitons, that is $ka\ll 1$. For this we can use the parabolic approximation for the dispersion of the excitons, $\omega_{ex}(k)=\omega_{ex}(0)+\frac{\hbar k^2}{2m_{ex}}$, where $m_{ex}$ is the exciton effective mass. For isotropic atoms in a square lattice of cubic symmetry, and in including only nearest neighbor interactions with coupling parameter $J$, we get $m_{ex}=-\hbar/(2Ja^2)$, and $\omega_{ex}(0)=\omega_A-2J$. For a large enough $2D$ optical lattice, the sum over ${\bf k'}$ can be approximated by a $2D$ integral given by 
\begin{equation} 
\frac{1}{N}\sum_{\bf k'}\rightarrow \left(\frac{a}{2\pi}\right)^2\int d^2k'. 
\end{equation} 
We get the two integrals 
\begin{eqnarray} \label{INT} 
I_{os}&=&-\frac{2m_{ex}}{\hbar^2}\left(\frac{a}{2\pi}\right)^2\int d^2k'\lim_{\eta\rightarrow 0_+}\frac{e^{i{\bf k'}\cdot{\bf r}_i}}{k^2-k'^2+i\eta}, \nonumber \\ 
I_{st}&=&-\frac{2m_{ex}}{\hbar^2}\left(\frac{a}{2\pi}\right)^2\int d^2k'\lim_{\eta\rightarrow 0_+}\frac{1}{k^2-k'^2+i\eta}. 
\end{eqnarray} 
as shown in the appendix. The scattered exciton, following the definition in \cite{Petrov}, is given by 
\begin{equation} 
\psi_i=\frac{1}{\sqrt{N}}\left\{e^{i{\bf k}\cdot{\bf r}_i}-f(k)\sqrt{\frac{i\pi}{2kr}}e^{ikr}\right\}, 
\end{equation} 
where $r=|{\bf r}_i|$ is the distance of the $i$ site from the origin. We defined the scattering amplitude by 
\begin{equation} 
f(k)=\frac{\frac{\pi E_a}{2\Delta_{ex}}}{1+\frac{\pi E_a}{2\Delta_{ex}}\left[\ln\left(\frac{ka}{\pi}\right)-i\frac{\pi}{2}\right]}, 
\end{equation} 
and the effective exciton band width is defined by 
\begin{equation} \label{EXBAND} 
\Delta_{ex}=\frac{\hbar^2\pi^2}{2m_{ex}a^2}. 
\end{equation} 
 
The dipole-dipole interaction energy between different sites in optical lattices is small, and hence the excitons have a small band width. In our case we have the limit of $ E_a\gg\Delta_{ex}$ so that the scattering amplitude is $f(k)\approx\frac{1}{\ln\left(\frac{ka}{\pi}\right)}$, which exactly reproduces the result for the scattering off a hard disk of radius $a$ \cite{Lapidus}, with the scattering cross section defined by $\sigma(k)=2\pi|f(k)|^2$. 
 
We thus conclude that an impurity generated by a missing atom in an optical lattice acts effectively like a hard disk of radius $a$. As the scattering is elastic, for an incident exciton with a fixed and small wave vector ${\bf k}$ we get a scattered exciton with a wave vector ${\bf k'}$, which equals in magnitude to the incident one, namely $|{\bf k}|=|{\bf k'}|=k$. Hence if a big number of incident excitons with identical wave vector are scattered off the impurity, we get a ring of radius $k$ of scattered excitons. 
 
The scattering amplitude is plotted in figure (2) as a function of wave vectors, $k$, for lattice constant $a=2000\ \AA$. The singularity at $k=0$ is the $2D$ signature. 
 
\begin{figure}[h!] 
\centerline{\epsfxsize=8.0cm \epsfbox{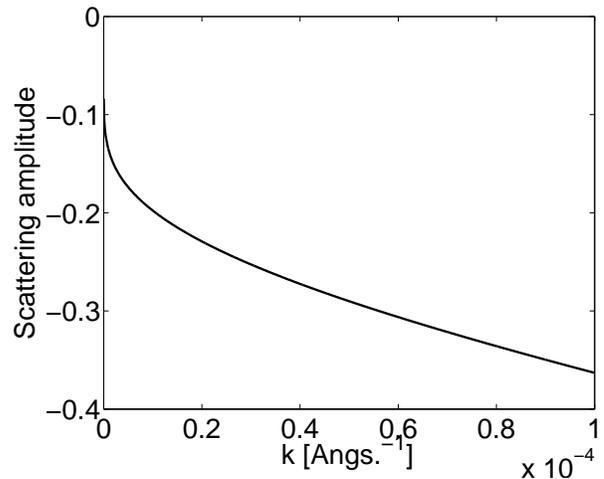}} 
\caption{The scattering amplitude vs. wave vector $k$, for zero detuning.} 
\end{figure} 
 
\section{Cavity polariton scattering of a vacancy in an optical lattice} 
 
Let us now add cavity mirrors to our lattice as described in \cite{ZoubiA}, where in the strong coupling regime, the system eigenstates are cavity polaritons. As polariton is a coherent superposition of an exciton and a photon it can scatter of the impurity due its excitonic part. The corresponding Hamiltonians then read: 
\begin{equation} 
H_0=\sum_{{\bf k}r}\hbar\Omega_r(k)\ A_{{\bf k}r}^{\dagger}A_{{\bf k}r},\ V=-E_a\ B_0^{\dagger}B_0, 
\end{equation} 
where $\Omega_{\pm}(k)$ are the upper and lower polariton branch dispersions, given by: $\Omega_{\pm}(k)=\frac{\omega_{cav}(k)+\omega_{ex}(k)}{2}\pm\Delta_k$. Here $\Delta_k=\sqrt{\delta_k^2+|g_k|^2}$, and the exciton-photon detuning is defined by $\delta_k=\frac{\omega_{cav}(k)-\omega_{ex}(k)}{2}$. The parameter $g_{\bf k}$ is for the exciton-photon coupling, which is taken to be of the electric dipole interaction. The cavity-photon dispersion is given by $\omega_{cav}(k)=\frac{c}{\sqrt{\epsilon}}\sqrt{k^2+\left(\frac{\pi}{L}\right)^2}$, where $L$ is the distance between the cavity mirrors, and $\epsilon$ is the cavity medium dielectric constant, which here taken to be a vacuum with $\epsilon=1$. The polariton operators are defined by $A_{{\bf k}\pm}=X_k^{\pm}\ B_{\bf k}+Y_k^{\pm}\ a_{\bf k}$, with $a_{\bf k}$ the cavity photon operator. $X_k^{\pm}$ and $Y_k^{\pm}$ are the exciton and photon amplitude, respectively, which are given by $X_k^{\pm}=\pm\sqrt{\frac{\Delta_k\mp\delta_k}{2\Delta_k}}$, and $Y_k^{\pm}=\frac{g_k}{2\Delta_k(\Delta_k\mp\delta_k)}$ (more details are found in Ref.\cite{ZoubiA}). 
 
As next step we now calculate the polariton scattering off such impurity. Again the scattering is elastic so that an incident polariton in branch $r$ with wave vector ${\bf k}$ will scatter into a polariton with wave vector ${\bf k'}$ in the same branch. 
 
The free eigenstates obey $H_0|\phi^{pol}_{{\bf k}r}\rangle=E_r(k)|\phi^{pol}_{{\bf k}r}\rangle$, and the whole system eigenstates obey $H|\psi\rangle=E|\psi\rangle$, where $E_r(k)=\hbar\Omega_r(k)$. The cavity photon eigenstate is defined by $a_{\bf k}^{\dagger}|vac\rangle=|\phi_{\bf k}^{cav}\rangle$, and the exciton eigenstate in quasi-momentum space is defined by $B_{\bf k}^{\dagger}|vac\rangle=|\phi_{\bf k}^{ex}\rangle$. The $r$ polariton branch eigenstate is defined by $A_{{\bf k}r}^{\dagger}|vac\rangle=|\phi_{{\bf k}r}^{pol}\rangle$, and in terms of exciton and photon states we get $|\phi_{{\bf k}\pm}^{pol}\rangle=X_k^{\pm}\ |\phi_{\bf k}^{ex}\rangle+Y_k^{\pm}\ |\phi_{\bf k}^{cav}\rangle$. In terms of real lattice space, we get $|\phi_{{\bf k}\pm}^{pol}\rangle=\frac{X_k^{\pm}}{\sqrt{N}}\sum_ie^{i{\bf k}\cdot{\bf r}_i}\ |\phi_i^{ex}\rangle+Y_k^{\pm}\ |\phi_{\bf k}^{cav}\rangle$. A general state $|\psi\rangle$ can be expanded as $|\psi\rangle=\sum_{\bf k}\psi_{\bf k}^{cav}\ |\phi_{\bf k}^{cav}\rangle+\sum_i\psi_i^{ex}\ |\phi_i^{ex}\rangle$. 
 
As above the scattered polariton state is given by the Schwinger-Lippmann equation $|\psi\rangle=|\phi^{pol}_{{\bf k}r}\rangle+G_0V|\psi\rangle$, where we used the free Green function operator in Eq.(\ref{Green}). We multiply the equation from the left by the lattice exciton eigenstate $\langle\phi_i^{ex}|$, and insert the previous identity operator between $G_0$ and $V$, to get 
\begin{equation} 
\langle\phi_i^{ex}|\psi\rangle=\langle\phi_i^{ex}|\phi^{pol}_{{\bf k}r}\rangle+\sum_j\langle\phi_i^{ex}|G_0|\phi_j^{ex}\rangle\langle\phi_j^{ex}|V|\psi\rangle. 
\end{equation} 
Using the above definitions, we get the scattered polariton excitonic part amplitude by 
\begin{equation} 
\psi_i^{ex}=\frac{X_k^r}{\sqrt{N}}e^{i{\bf k}\cdot{\bf r}_i}-\frac{E_a}{N}\sum_{{\bf k'}s}\lim_{\eta\rightarrow 0_+}\frac{|X_{k'}^s|^2\ e^{i{\bf k'}\cdot{\bf r}_i}}{E_r(k)-E_s(k')+i\eta}\ \psi_0^{ex}, 
\end{equation} 
where due to energy conservation we replaced the energy $E$ by the initial energy $E_r(k)$, and $|X_{k'}^s|^2$ is the exciton weight in the polariton. For the impurity site amplitude we have 
\begin{equation} 
\psi_0^{ex}=\frac{X_k^r}{\sqrt{N}}\left\{1+\frac{E_a}{N}\sum_{{\bf k'}s}\lim_{\eta\rightarrow 0_+}\frac{|X_{k'}^s|^2}{E_r(k)-E_s(k')+i\eta}\right\}^{-1}. 
\end{equation} 
We thus obtain 
\begin{equation} 
\psi_i^{ex}=\frac{X_k^r}{\sqrt{N}}\left\{e^{i{\bf k}\cdot{\bf r}_i}+\frac{E_aI_{os}}{1-E_aI_{st}}\right\}, 
\end{equation} 
where we defined oscillating and static sums. For a large optical lattice size the $k$-space can be assumed continuous and the sums converted to the integrals 
\begin{eqnarray} 
I_{os}&=&-\sum_s\left(\frac{a}{2\pi}\right)^2\int d^2k'\lim_{\eta\rightarrow 0_+}\frac{|X_{k'}^s|^2\ e^{i{\bf k'}\cdot{\bf r}_i}}{E_r(k)-E_s(k')+i\eta}, \nonumber \\ 
I_{st}&=&-\sum_s\left(\frac{a}{2\pi}\right)^2\int d^2k'\lim_{\eta\rightarrow 0_+}\frac{|X_{k'}^s|^2}{E_r(k)-E_s(k')+i\eta}. \nonumber \\ 
\end{eqnarray} 
 
We calculate first the oscillating integral in case of scattering of small wave vector polaritons, that is in the limit $ka\ll1$ in the lower branch. In this limit the polariton dispersion can considered approximately parabolic with a polariton effective mass of $m_p$. This is of the order of the cavity photon effective mass $m_p\approx (\hbar\pi)/(cL)$. Hence, the lower polariton branch dispersion is taken to be $E_-(k)=E_-(0)+\frac{\hbar^2k^2}{2m_p}$. The scattered states will also have small wave vectors due to the energy conservation. As the upper branch has higher energies its contributions are negligibly small here. Anyway the upper branch for larger wave vectors is mainly photonic with small excitonic part and thus only weakly contributes to the scattering of impurities. We consider the case around zero detuning between the excitons and photons. Therefore the excitonic weight $|X_{k}^-|^2$ will change from half around zero wave vector up to one for large wave vectors, where the lower branch became excitonic. The weight $|X_{k}^-|^2$ is a smooth function of $k$, and in the present case can be taken out of the integral, and to be fixed with the initial polariton excitonic weight value. From here on we neglect the contribution of the upper polariton branch, both as initial and scattered states, and also as intermediate scattering state. We will drop the branch index and summation, and all the parameters will be only for the lower branch. 
 
The integral now reads: 
\begin{eqnarray} 
I_{os}=-X_{k}^2\ \frac{2m_p}{\hbar^2}\left(\frac{a}{2\pi}\right)^2\int d^2k'\lim_{\eta\rightarrow 0_+}\frac{e^{i{\bf k'}\cdot{\bf r}_i}}{k^2-k^{'2}+i\eta}. 
\end{eqnarray} 
This is similar to the case of excitons of the previous section except for the factor $|X_{k}^r|^2$ and involving the much smaller polariton effective mass in place of the exciton mass. Hence we have $m_p\ll m_{ex}$ and the integration (see the appendix) gives 
\begin{equation} 
I_{os}=-\frac{\pi X_k^2}{2\Delta_p}\sqrt{\frac{i\pi}{2kr}}e^{ikr}, 
\end{equation} 
where we defined the polariton effective band width $\Delta_p=\frac{\hbar^2\pi^2}{2m_{p}a^2}$. 
 
We turn now to the second static integral. Here no oscillating exponent appears, and hence all wave vectors along the lower branch contribute, so that the intermediate scattering state can be anywhere along the lower branch.  Again for energy reasons we neglect the contribution of the upper branch. The integral then can be simplified by using a model for the lower branch dispersion in place of the real one \cite{ZoubiC}. The lower branch is divided into two parts. The first part between $0\leq k\leq k_0$ is taken to be of a parabolic dispersion with a polariton effective mass, where $E(k)=E(0)+(\hbar^2k^2)/(2m_p)$; and the second part between $k_0\leq k\leq \pi/a$, where $\pi/a$ is the Brillouin boundary, is taken to be  dispersion less with energy equal to the exciton energy at zero wave vector, where we have $k_0\ll\pi/a$. The intersection point between the two parts is fixed by $\hbar^2k_0^2/(2m_p)=E_0-E(0)$. The integration as shown in the appendix gives 
\begin{equation} \label{ST} 
I_{st}=-\frac{\pi X_k^2}{2\Delta_p}\left[\ln\left(\frac{k}{k_0}\right)-i\frac{\pi}{2}\right]+\frac{\pi}{4\Lambda_k}, 
\end{equation} 
where $\Lambda_k=E_0-E(k)$. The first term is the contribution of the parabolic part and the second originates from the flat part. The $Ln(k/k_0)$ is large in the limit $ka\ll 1$, but for a laterally confined $2D$ optical lattice, and for the smallest wave vector $k_m$ which is different from zero, due to the fact $\Delta_p\gg E_a$, the term $\frac{\pi X_{k_m}^2}{2\Delta_p}\ln\left(\frac{k_m}{k_0}\right)$ is much smaller than one and can be neglected. Hence $I_{st}\approx\pi/(4\Lambda_k)$. 
 
The scattered polariton excitonic part reads 
\begin{equation} 
\psi_i^{ex}=\frac{X_k}{\sqrt{N}}\left\{e^{i{\bf k}\cdot{\bf r}_i}-f(k)\sqrt{\frac{i\pi}{2kr}}e^{ikr}\right\}, 
\end{equation} 
where the scattering amplitude is given by 
\begin{equation} 
f(k)=\frac{X_k^2\left(\frac{\pi E_a}{2\Delta_p}\right)}{1-\left(\frac{\pi E_a}{4\Lambda_k}\right)}. 
\end{equation} 
As $E_a>\Lambda_k$, we have $(\pi E_a)/(4\Lambda_k)>1$, and the scattering amplitude is negative, that is $f(k)<0$. Again the above impurity effective potential is repulsive for the polaritons. 
 
In the case of zero detuning, that is $\delta_0=0$, we have for small wave vectors $X_k^2= 1/2$, and $\Lambda_k$ is of the order of the exciton-photon coupling, $|g_k|$, where $E_a\gg\Lambda_k$, and we get $f(k)\approx-\Lambda_k/\Delta_p$. This scattering amplitude equals that of the scattering from an effective potential of a square barrier potential of height $\Lambda_k$ and width $a$.  
 
For the case of negative detuning, in the limit of $E_a\gg\delta_0$, where $\Lambda_k$ is of the order of the exciton-photon detuning, the scattering amplitude is $f(k)\approx-(2X_k^2\Lambda_k)/\Delta_p$. This scattering amplitude equals that of the scattering from an effective potential of a square barrier potential of height $2X_k^2\Lambda_k$ and width $a$. Hence we find that the scattering amplitude can be controlled by changing the exciton-photon detuning. 
 
Lets insert some numbers: for an atomic energy $E_a=2\ eV$, exciton-photon coupling $\hbar|g|=0.0001\ eV$, and lattice constant $a=2000\ \AA$, the scattering amplitude is plotted in figure (3) as a function of the exciton-photon detuning, for approximately normal incident waves, where we used $k=10^{-6}\ \AA$. It is clear that maximum scattering is obtained for zero detuning. As the detuning increases, both positive or negative, the $k\approx 0$ polaritons become more photonic and they scattered off the impurity much less. As the atomic energy $E_a$ is much larger than the coupling energy $\hbar g$, the scattering amplitude only shows a negligibly small asymmetry for negative and positive detunings, but we expect a clear resonance of the scattering amplitude around resonance $\delta_0$. Considering a stream of incident polaritons with the same wave vector ${\bf k}=k\hat{\bf k}$ we then should see a ring of radius $k$ of scattered polaritons. 
 
\begin{figure}[h!] 
\centerline{\epsfxsize=8.0cm \epsfbox{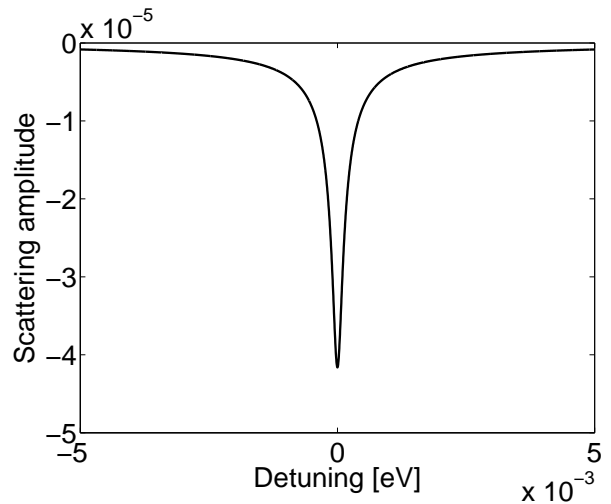}} 
\caption{The scattering amplitude vs. exciton-photon detuning, for $k\approx 0$ polaritons.} 
\end{figure} 
 
\section{Exciton and polariton scattering of a vacancy in two-atoms per site optical lattices} 
 
We will now go one step further and study the scattering of excitons and polaritons in a doubly occupied optical lattice where one atom is missing in one site (called the origin site). The corresponding exciton polariton picture for an ideal lattice is discussed in Ref.\cite{ZoubiB}. Here resonant electronic excitation transfer occurs among the two on-site atoms with transfer parameter $J_0$ and also among the nearest neighbor sites with transfer parameter $J_1$. The on-site interaction can be diagonalized to form symmetric and antisymmetric localized excitation with frequencies $\omega_{s}=\omega_A+J_0$, and $\omega_{a}=\omega_A-J_0$, respectively. Only the symmetric states have a significant dipole moment and thus form excitons through resonant coupling. The symmetric exciton dispersion in the limit $ka\ll1$ then reads $\omega_s(k)=\omega_A+J_0+8J_1+\frac{\hbar k^2}{2m_{eff}}$, with the effective mass $m_{eff}=-\hbar/(4J_1a^2)$ and up to a rescaling is analogous to the previous case. Naturally within a cavity only the symmetric excitons are coupled to the photons to form cavity polaritons (for details see \cite{ZoubiB}). At first sight there seems to be nothing new in this case. However, we will show that the defect, plotted schematically in figure (4), which in this case constitutes a single atom as compared to two, here acts substantially different from the simple no atom hole in the case of a single occupied lattice. Physically this can be understood from the fact that a single atom can store almost the same amount of energy as a pair if only one photon excitations are considered.  
 
Let us now calculate the scattering amplitude of excitons and polaritons of this impurity. The scattering is for the symmetric excitons, and for the polariton excitonic part. The antisymmetric states are not involved in the scattering process, as they are on-site localized and posses no dipole moment. 
 
\begin{figure}[h!] 
\centerline{\epsfxsize=8.0cm \epsfbox{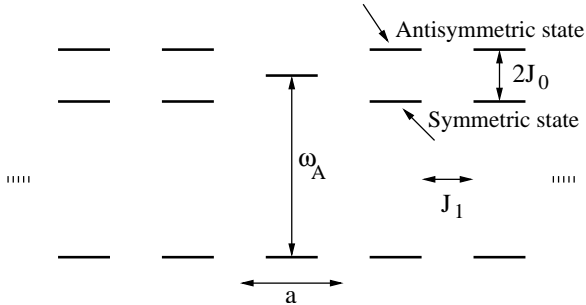}} 
\caption{An impurity in $1D$ optical lattice.} 
\end{figure} 
 
Again we concentrate on the scattering of long wave length excitons with a parabolic dispersion described by an effective mass. The scattering amplitude is found to be 
\begin{equation} 
f(k)=\frac{\frac{\pi J_0}{2\Delta_{ex}}}{1+\frac{\pi J_0}{2\Delta_{ex}}\left[\ln\left(\frac{ka}{\pi}\right)-i\frac{\pi}{2}\right]}. 
\end{equation} 
For negative on-site energy transfer, $J_0<0$, the scattering is of an effective potential of a square barrier of height $|J_0|$ and width $a$. One need to compare between the on-site transfer parameter $|J_0|$ and the exciton band width $\Delta_{ex}$. In the limit of $|J_0|\gg\Delta_{ex}$ the scattering is identical to that from an effective potential of a hard disk with radius $a$ (see figure (2)), that is $f(k)\approx\frac{1}{\ln\left(\frac{ka}{\pi}\right)}$. 
 
For polaritons the scattering of the impurity is only for their excitonic part. We consider the scattering of long wave length lower branch polaritons with a parabolic dispersion and an effective polariton mass $m_p$. The scattering amplitude is given by 
\begin{equation} 
f(k)=\frac{X_k^2\left(\frac{\pi J_0}{2\Delta_p}\right)}{1-\left(\frac{\pi J_0}{4\Lambda_k}\right)}. 
\end{equation} 
Here $\Lambda_k=E_0-E(k)$, where $E_0=\hbar\omega_A+\hbar J_0$, and where $J_0\gg J_1$. 
 
In the limit of $|J_0|\gg\Lambda_k$ we get $f(k)\approx -X_k^2\left(\frac{2\Lambda_k}{\Delta_p}\right)$, where, for $J_0<0$, the scattering acts like a square barrier of height $2X_k^2\Lambda_k$ and width $a$. Similarly in the opposite limit of $|J_0|\ll\Lambda_k$ we get $f(k)\approx X_k^2\left(\frac{\pi J_0}{2\Delta_p}\right)$, and the scattering, for $J_0<0$ mimics a square barrier of height $\pi |J_0|X_k^2/2$ and width $a$. As shown in the previous chapter, the polariton scattering of an impurity can be modulated by controlling the exciton-photon detuning. As we change $\delta_k$, then $X_k^2$ and $\Lambda_k$ are strongly changed, and as a result the scattering amplitude $f(k)$ and the effective potential strength are changed. 
 
Introducing the numerical examples used in the previous section with transfer energy $\hbar J_0=-0.001\ eV$ the scattering amplitude is plotted in figure (5) as a function of the exciton-photon detuning, for incident waves, namely $k\approx 0$. It is clear that maximum scattering is obtained for zero detuning. As here the transfer energy $\hbar J_0$ is smaller than, and close to, the coupling energy $\hbar g$, the scattering amplitude asymmetry for negative and positive detuning is much more pronounced relative the case of the previous section. For $k=0$, at zero detuning, the lower branch polariton is half exciton and half photon. For negative detuning, at $k=0$, the lower branch polariton is more photonic than excitonic; while for positive detuning the opposite the lower branch polariton is more excitonic than photonic. Therefore, the positive detuning scattering amplitude is larger than the negative one. 
 
\begin{figure}[h!] 
\centerline{\epsfxsize=8.0cm \epsfbox{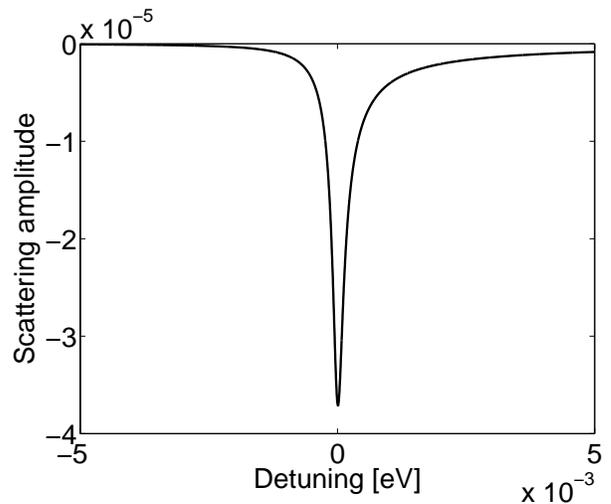}} 
\caption{The scattering amplitude vs. exciton-photon detuning, for $k\approx 0$ polaritons.} 
\end{figure} 
 
\subsection{An impurity in an asymmetric optical lattice} 
 
One of the most interesting phenomena for excitons in doubly occupied lattices appears in the case of asymmetric optical lattice sites \cite{ZoubiB}. We get a strong polarization dependence of the exciton and polariton energies. Here we discuss   
the consequence of this for their scattering amplitudes off impurities. 
 
If one of the orthogonal pairs of counter propagating lasers forming the lattice potentials has a different intensity, the potential is elongated in one direction, e.g. $x$. Hence the two atoms at one site will have a larger average distance $R$ in the $x$ direction as shown in figure (6). The atomic transition dipole induced by the cavity photon is assumed to be in the (x,y)-plane $\vec{\mu}=\mu(\cos\theta,\sin\theta)$, where $\theta$ is the angle between the dipole $\vec{\mu}$ and the $x$ axis. The resonance dipole-dipole interaction between the two on-site atoms is $\hbar J_0(\theta)=\hbar \bar{J}\ \left(1-3\cos^2\theta\right)$, where $\hbar \bar{J}=\frac{\mu^2}{4\pi\epsilon_0R^3}$. We have $\hbar J_0(\theta=0)=-2\hbar \bar{J}$, and $\hbar J_0(\theta=90)=\hbar \bar{J}$, with $\hbar J_0(\theta\approx 54.74)=0$. The detuning energy between symmetric and antisymmetric states thus will change sign at some angle, where the polarization even vanishes.  
 
\begin{figure}[h!] 
\centerline{\epsfxsize=5.0cm \epsfbox{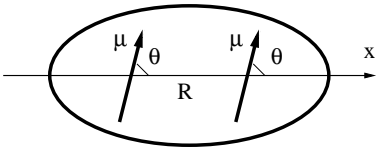}} 
\caption{An asymmetric optical site.} 
\end{figure} 
 
The long wave length polariton scattering amplitude as a function of the relative transition dipole direction now is 
\begin{equation} \label{SSS} 
f_k(\theta)=\frac{X_k^2(\theta)\left(\frac{\pi J_0(\theta)}{2\Delta_p}\right)}{1-\left(\frac{\pi J_0(\theta)}{4\Lambda_k(\theta)}\right)}. 
\end{equation} 
We plot this scattering amplitude as a function of $\theta$ for different values of $\bar{J}$, and for $k\approx 0$, again using the numerical values of section 3. We also have $E_c(0)=E_A+J(0)$, and $\hbar J_1=10^{-7}\ eV$. In figure (7) the plot is for $\hbar\bar{J}=10^{-4}\ eV$, where $\hbar|g|=\hbar\bar{J}$. Figure (8) is for $\hbar\bar{J}=5\times 10^{-4}\ eV$, where $\hbar|g|<\hbar\bar{J}$. 
Note that the scattering amplitude changes sign at $\theta\approx 54.74$ and we can turn of scattering by using this proper angle. In figure (9) we plot this for $\hbar\bar{J}=10^{-3}\ eV$. Here two resonances appear. Figure (10) is for $\hbar\bar{J}=5\times 10^{-3}\ eV$. Now the resonances tend to the negative-positive crossover angle. 
 
The negative scattering amplitude corresponds to a repulsive effective potential, and polaritons scattered away of the impurity. While the positive scattering amplitude corresponds to an attractive potential. Here even the formation of bound states could be expected, allowing for localized polaritons. The bound state signature appears in the pole of the scattering amplitude of Eq.(\ref{SSS}) but by including the $Ln$ term of Eq.(\ref{ST}) in the denominator. We get a shallow bound state that falls inside the polariton line width, and which presents as a resonance scattering state inside a continuum. 
 
\begin{figure}[h!] 
\centerline{\epsfxsize=8.0cm \epsfbox{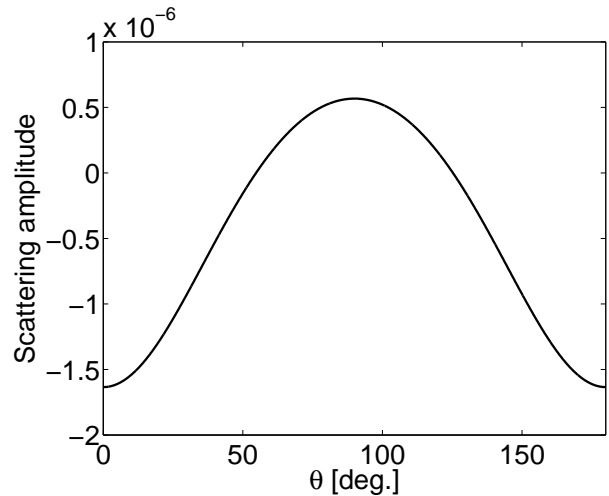}} 
\caption{The scattering amplitude vs. $\theta$, for $k\approx 0$ polaritons, with $\hbar\bar{J}=10^{-4}\ eV$.} 
\end{figure} 
 
\begin{figure}[h!] 
\centerline{\epsfxsize=8.0cm \epsfbox{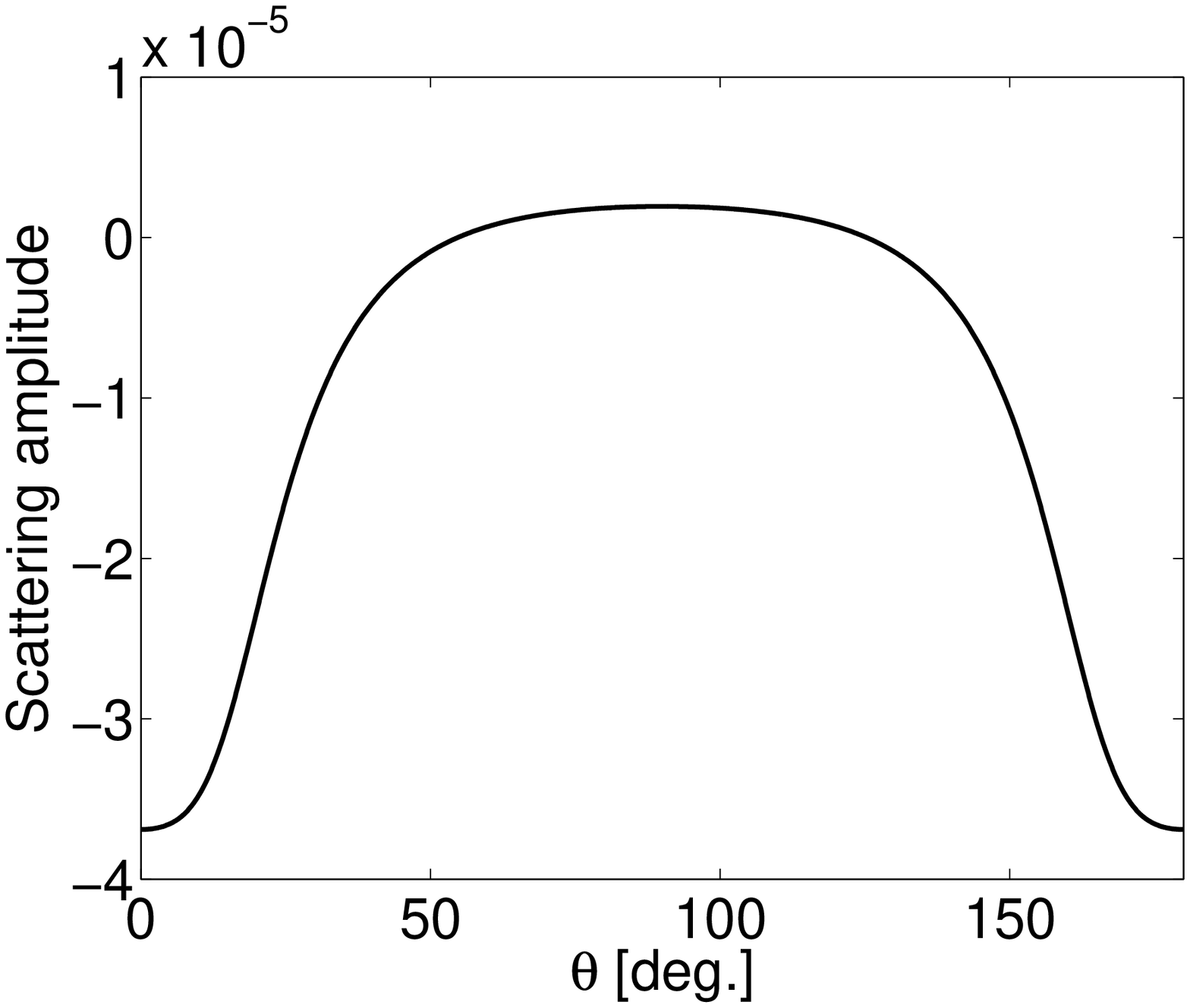}} 
\caption{The scattering amplitude vs. $\theta$, for $k\approx 0$ polaritons, with $\hbar\bar{J}=5\times 10^{-4}\ eV$.} 
\end{figure} 
 
\begin{figure}[h!] 
\centerline{\epsfxsize=8.0cm \epsfbox{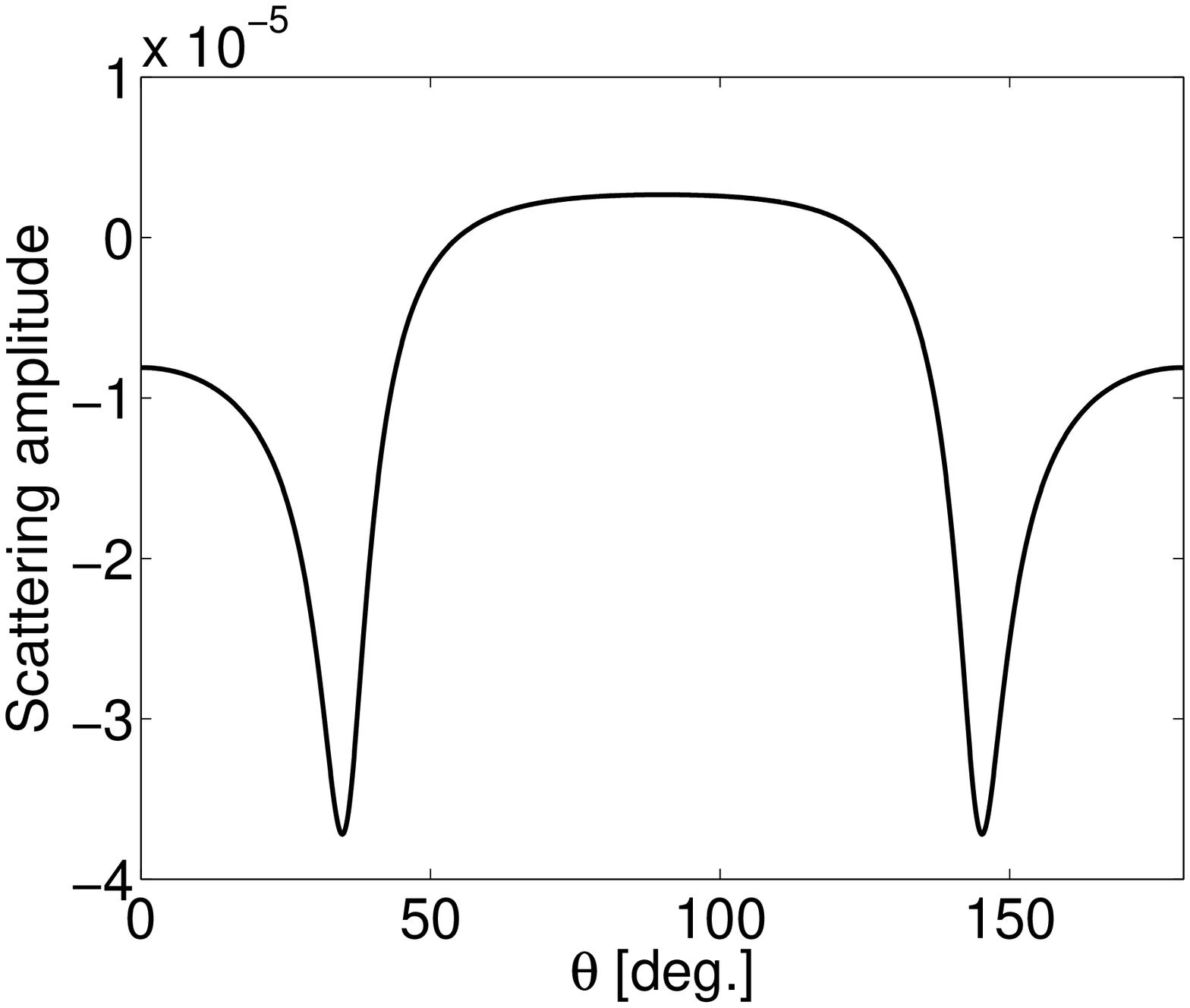}} 
\caption{The scattering amplitude vs. $\theta$, for $k\approx 0$ polaritons, with $\hbar\bar{J}=10^{-3}\ eV$.} 
\end{figure} 
 
\begin{figure}[h!] 
\centerline{\epsfxsize=8.0cm \epsfbox{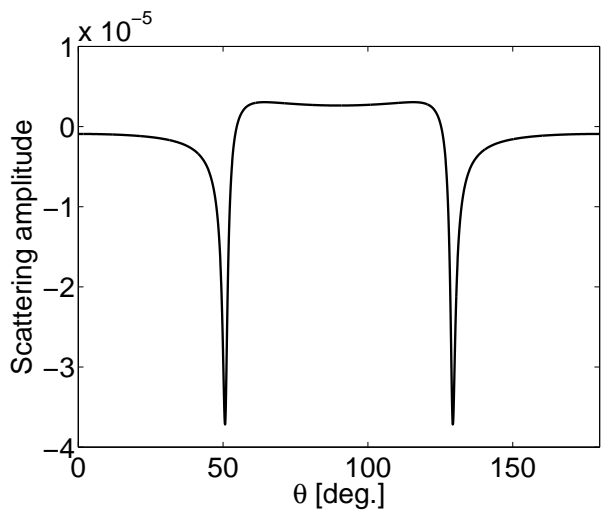}} 
\caption{The scattering amplitude vs. $\theta$, for $k\approx 0$ polaritons, with $\hbar\bar{J}=5\times 10^{-3}\ eV$.} 
\end{figure} 
 
\section{Summary} 
 
We demonstrated that defects in an optical lattice in the Mott insulator phase will change the dynamics of resonant excitons and cavity polaritons. For a very low density of defects, the exciton and polariton pictures still hold and the defect simply acts as a scatterer for such quasi-particles. We calculated the scattering amplitude for excitons and polaritons showing that the defects can be approximated by effective potentials. In the case of one atom per site the vacancy scattering effective potential for excitons is a hard disk of a radius equal to the lattice constant $a$, which is quite different for polaritons, where only the excitonic part contributes significantly to the scattering. The scattering effective potential for long wave length polaritons just is a square barrier of height $2|X_k|^2\Lambda_k$ and width $a$, where $|X_k|^2$ is the exciton weight in the cavity polariton, and which appears here as the scattering is only for the polariton excitonic part. For the case of zero exciton-photon detuning, and in the limit $ka\ll1$, we have $|X_k|^2=1/2$, and $\Lambda_k$ is of the order of the Rabi splitting. While for large negative detuning we have $|X_k|^2\ll1$, and $\Lambda_k$ is of the order of the exciton-photon detuning. As $|X_k|^2$ and $\Lambda_k$ are a function of the detuning, hence the scattering amplitude can be controlled by changing the detuning, with a scattering resonance at zero detuning. 
 
In the case of a two atoms per site lattice with a defect represented by a single atom site, we found two distinct cases. If the on site dipole-dipole coupling $J_0$ is larger than the Rabi splitting, we got similar results as in the case of one atom per site. But if $J_0$ is smaller than the Rabi splitting, hence the polariton scattering effective potential is identical to the scattering of a square barrier of height $J_0$ and width $a$. The parameter $J_0$ can be easily controlled in changing the photon polarization, as we presented for the case of asymmetric optical lattice sites. Also we showed that at a fixed angle $J_0$ can change sign from positive to negative, and then also the scattering amplitude changes sign. 
 
We conclude that cavity polaritons can be used as a useful tool to observe defects in an optical lattice. As the polariton is part exciton and part photon, where the photon part can be controlled externally, the above results suggest that one can recognize the kind of defect in each case. For example in transmission or reflection experiments, for an incident field with a fixed wave vector $k$, in the transmitted or reflected signals we get a scattering ring of radius $k$ with the appropriate intensity which obtained from the scattering cross section.  
 
\begin{acknowledgments} 
The work was supported by the Austrian Science Fund (FWF), through the Lise-Meitner Program (M977).  
\end{acknowledgments} 
 
\appendix  
 
\section{The calculation of Eq.(\ref{INT}) integrals} 
 
Here we give the calculation details of the two integrals of Eqs.(\ref{INT}). 
 
The oscillating integral is written explicitly as 
\begin{equation} 
I_{os}=-\frac{2m_{ex}}{\hbar^2}\left(\frac{a}{2\pi}\right)^2\int_0^{+\infty} dk'\int_{-\pi}^{+\pi}d\theta\lim_{\eta\rightarrow 0_+}\frac{k'e^{ik'r\sin\theta}}{k^2-k'^2+i\eta}, 
\end{equation} 
where $r=|{\bf r}_i|$ is the distance of the $i$ site from the origin, and $k'=|{\bf k'}|$. The integral over $k'$ is from zero to infinity, which is possible here due to the oscillating exponents, where the contributions of large wave vectors are negligible. The integral over $\theta$ is calculated by 
\begin{equation} 
J_0(k'r)=\frac{1}{2\pi}\int_{-\pi}^{+\pi}d\theta\ e^{ik'r\sin\theta}, 
\end{equation} 
where $J_0(x)$ is the zero order Bessel function of the first kind. We have now 
\begin{equation} 
I_{os}=\frac{2m_{ex}}{\hbar^2}\left(\frac{a}{2\pi}\right)^22\pi\int_0^{+\infty} dk'\lim_{\eta\rightarrow 0_+}\frac{k'J_0(k'r)}{k'^2-k^2-i\eta}. 
\end{equation} 
We use the result 
\begin{equation} 
\int_0^{+\infty} dk'\lim_{\eta\rightarrow 0_+}\frac{k'J_0(k'r)}{k'^2-k^2-i\eta}=i\frac{\pi}{2}H_0^{(1)}(kr), 
\end{equation} 
where $H_0^{(1)}(x)$ is the zero order of the Hankel function of the first kind, which is defined by $H_0^{(1)}(x)=J_0(x)+iN_0(x)$, where $N_0(x)$ is the zero order Bessel function of the second kind. The scattered wave is observed very far from the impurity, that is at $r\rightarrow +\infty$. For large $x$ we have the asymptotic expansions 
\begin{equation} 
J_0(x)\sim\sqrt{\frac{2}{\pi x}}\cos\left(x-\frac{\pi}{4}\right)\ ,\ N_0(x)\sim\sqrt{\frac{2}{\pi x}}\sin\left(x-\frac{\pi}{4}\right), 
\end{equation} 
which yield 
\begin{equation} 
H_0^{(1)}(x)\sim\sqrt{\frac{2}{\pi x}}\ e^{i(x-\pi/4)}. 
\end{equation} 
The result is 
\begin{equation} 
I_{os}=-\frac{\pi}{2\Delta_{ex}}\sqrt{\frac{i\pi}{2kr}}e^{ikr}, 
\end{equation} 
where we defined the effective exciton band width in Eq.(\ref{EXBAND}). 
 
For the static integral the angle integral gives $2\pi$, we have 
\begin{equation} 
I_{st}=\frac{2m_{ex}}{\hbar^2}\left(\frac{a}{2\pi}\right)^22\pi\int_0^{\pi/a} dk'\lim_{\eta\rightarrow 0_+}\frac{k'}{k'^2-k^2-i\eta}. 
\end{equation} 
Here, due to the fact that no oscillations exist, it is not possible to extend the integral to infinity. We used a cut-off at the Brillouin zone boundary. Such cut-off is a physical one as the maximum exciton wave vector is for wave length equals the lattice constant. The integration, in the limit of $ka\ll1$, yields 
\begin{equation} 
I_{st}=-\frac{\pi}{2\Delta_{ex}}\left[\ln\left(\frac{ka}{\pi}\right)-i\frac{\pi}{2}\right]. 
\end{equation}

\end{document}